\documentclass[useAMS,usenatbib]{mn2e}

\usepackage{graphicx,lscape}

\title[Stellar populations in NGC59]{The stellar populations in the low luminosity, early-type galaxy NGC59}
\author[A.E. Sansom et al.]
{A. E. Sansom,$^{1}$\thanks{E-mail: aesansom@uclan.ac.uk} 
J.~J.~Thirlwall,$^{1}$ 
M.~A.~Deakin,$^{1}$
P.~V\"ais\"anen,$^{2,3}$
A.~Y.~Kniazev,$^{2,3,4}$
\newauthor 
and J. Th. van Loon,$^{5}$ \\
$^{1}$Jeremiah Horrocks Institute, University of Central Lancashire, 
Preston PR1 2HE, UK\\
$^{2}$South African Astronomical Observatory, P.O. Box 9 Observatory, Cape Town, South Africa \\ 
$^{3}$Southern African Large Telescope, P.O. Box 9 Observatory, Cape Town, South Africa \\ 
$^{4}$Sternberg Astronomical Institute, Lomonosov Moscow State University, Universitetskij Pr. 13, Moscow, 119992, Russia \\ 
$^{5}$Lennard-Jones Laboratories, Keele University, Staffordshire, ST5 5BG, UK \\ 
}
\begin{document}


\pagerange{\pageref{firstpage}--\pageref{lastpage}} \pubyear{2002}

\maketitle

\label{firstpage}

\begin{abstract}
Low luminosity galaxies may be the building blocks of more luminous systems.
Southern African Large Telescope (SALT) observations of the low luminosity,
early-type galaxy NGC59 are obtained and analysed. These data are used to 
measure the stellar population parameters in the centre and off-centre 
regions of this galaxy, in order to uncover its likely star formation history.
We find evidence of older stars, in addition to young stars in the emission
line regions. The metallicity of the stellar population is constrained to 
be [Z/H]$\sim-1.1$ to $-1.6$,
which is extremely low, even for this low luminosity galaxy, since it is not 
classed as a dwarf spheroidal galaxy. The 
measured [$\alpha$/Fe] ratio is sub-solar, which indicates an extended 
star formation history in NGC59. If such objects formed the building blocks
of more massive, early-type galaxies, then they must have been gaseous 
mergers, rather than dry mergers, in order to increase the metals to 
observed levels in luminous, early-type galaxies.
\end{abstract}



\section{Introduction}

The early-type galaxy NGC59 is similar in morphology to many other low 
luminosity E or S0 galaxies. It has a stellar mass of 
$\sim5\times10^8 M_{\odot}$. This is about 1/200th of M* (the knee in 
the overall galaxy mass function for galaxies in the local Universe), 
taking Log(M*/M$_{\odot}$) $=10.64$ from Kelvin et al. (2014).
These galaxies form an interesting class of objects whose place 
in the overall scheme of galaxy evolution has yet to be properly 
understood.

The hierarchical merger picture of galaxy evolution implies that lower mass 
galaxies formed before more massive galaxies and that they contributed to the 
content of more massive galaxies through mergers, over time. Early-type 
galaxies (ETGs, including elliptical and lenticular) are thought to be 
particularly affected by this process. Thus, some of the original lower 
mass galaxies would be consumed in producing giant ellipticals, whilst 
others, which did not take part in mergers, would still exist today. 
These ancient relics may look very different today from what they were 
like at earlier times. For example, at early times they must have contained 
gas, with which to make new stars, post-merging. This is because the 
stellar populations of giant galaxies are not the same as those in lower 
mass galaxies (e.g. Bender, Burstein \& Faber 1993; Worthey 1998). 
In particular, stars in giant galaxies tend to be more metal-rich than stars 
in lower mass galaxies. This is the well known mass-metallicity relation, 
which applies to all types of galaxies (e.g. Gallazzi et al. 2006; 
Foster et al. 2012). Massive ETGs are also known to have higher [Mg/Fe] 
ratios than lower mass ETGs (Worthey 1998), or more generally 
enhanced [$\alpha$/Fe] element ratios. These are generally interpreted 
as due to more rapid star formation timescales in massive ETGs, because 
of the relative timescales of different supernova contributions. However 
alternative explanations have been suggested for varying [$\alpha$/Fe] ratio, 
including systematic variations with galaxy mass of the initial stellar mass 
function or SNIa delay times, or selective wind mass losses may also affect 
this ratio (e.g. Pipino et al. 2009).

Lower mass galaxies that survive, without being destroyed in mergers, 
would continue with their star formation histories (SFHs), unabated. 
As yet there have not been many measurements or modelling of abundance 
patterns from integrated light in ETGs with velocity 
dispersions below about $\sigma<$80 kms$^{-1}$. Trends observed at higher 
velocity dispersions have been well studied (e.g. Pipino et al 2009; Graves, 
Faber \& Schiavon 2009; Conroy et al. 2014), showing hints of increased 
scatter at lower velocity dispersions. More diverse properties at lower mass
might be expected from the greater impact of star forming events on the
energetics of systems with smaller gravitational wells (Merlin et al. 
2012). 

With NTT data we previously investigated a small sample of nearby, low 
luminosity, early-type galaxies (hereafter LLEs) (Sansom \& Northeast 2008). 
The purpose was to compare their stellar populations and kinematics with 
their more luminous and more massive counterparts (closer to M* and above) 
and with similar mass bulges of spiral galaxies. We found that the LLEs fell 
below extrapolations of trends in metal line-strengths, but were consistent 
with spiral bulges of similar velocity dispersion. Fitted luminosity weighted 
mean ages were found to be young (Ages $<$3 Gyrs) and abundances covered a 
wide range ($-1.4<$[Fe/H]$<+0.1$), with low [$\alpha$/Fe] ratios (typically 
sub-solar). These stellar population characteristics are very different 
from those of luminous ETGs (e.g. Thomas et al. 2005). The only other galaxies 
where such characteristics have been detected are in some resolved star data 
for Local Group dwarf spheroidals (Venn et al. 2004, their fig.~2; 
Tolstoy et al. 2009), which are much fainter than our original sample of 
LLEs, and have stellar masses of only a few $\times 10^7 M_{\odot}$ or less 
($<$M*$/1000$).
The low [$\alpha$/Fe] results found for our sample of LLEs have been 
used to test initial mass functions (IMF) in galaxies, supporting a steeper 
integrated galactic IMF with increasing galaxy mass and decreasing star 
formation rate (Recchi, Calura and Kroupa 2009). LLEs provide important 
tests of scaling relations with mass and luminosity in general, since they 
are at one extreme of those relations.

A few other studies have measured detailed stellar population parameters in
LLEs, including that of Annibali et al. (2011), who studied a small sample 
of ETGs with M$_R$ in the range $-19.8$ to $-14.8$. They found generally 
sub-solar metallicities and sub-solar [$\alpha$/Fe] on average, with some 
hints of environmental dependancies, when comparing their ETGs in poor groups 
with results from the Coma cluster. Koleva et al. (2013) (and references 
therein)
measured stellar population age and [Fe/H] in samples of dwarf ETGs and 
transition-type dwarfs, covering M$_B=-19.0$ to $-14.5$ and found a wide 
range of, typically sub-solar metallicities, around $-0.7$ dex on average, 
from their luminosity weighted, simple stellar populations (SSP) fits.

The signal-to-noise in LLE spectra is generally not as high as for many 
luminous ETGs that have been the subject of stellar population analysis. 
Therefore results for LLEs may be more susceptible to the well known 
age-metallicity degeneracies that plague such analysis. It is for this 
reason that we wished to check our results for the most extreme LLEs, 
using new data that is more sensitive to stellar population parameters. 
From our NTT data we found that the S0 galaxy NGC59 (M$_B=-15.57$) had 
central stellar velocity dispersion of 37 $\pm$ 5 kms$^{-1}$ and the 
following stellar population parameter estimates: Age=1.5$\pm$0.1 Gyr, 
[Fe/H]$=-1.350 \pm 0.075$ and [$\alpha$/Fe]$=-0.300 \pm 0.075$ dex. Whilst 
other LLEs that we observed also had similarly unusual stellar populations, 
NGC59 was the most extreme. Therefore, in this paper we test the results 
for this galaxy with new data, taken with the SALT telescope (O'Donoghue 
et al. 2006), covering 
a wavelength range extending further into the blue region and using 
stellar population models, based on state-of-the-art stellar spectral 
libraries.

NGC59 is classified as SA0 in RC3 (de Vaucouleurs et al. 1991) 
and E in the APM Bright Galaxy Catalogue (Loveday 1996). It is usually 
classified as a dwarf, early-type galaxy. 
Bouchard et al. (2005) re-classified this galaxy to a dS0 Pec on the basis 
of its HI and H$\alpha$ detections.
Beaulieu et al. (2006) measured HI=1.4$\pm$0.1 $\times$ 10$^7$ M$\sun$ 
in NGC59 and suggest 
that this HI mass and m(HI)/L$_B$ ratio are closer to those of dIrr and 
normal spiral galaxies than dEs. Star formation activity was detected by 
Skillman, C\^{o}t\'{e} \& Miller (2003), in which the H$\alpha$ image 
shows lumpy, approximately central H$\alpha$ emission (their fig.~1). 
Images in several wavebands are also shown in Saviane et al. (2008, 
their fig.~1). The nucleus of NGC59 was resolved 
in J,H and K images, revealing two peaks $\approx$ 2.3 arcsec apart (de Swardt, 
Kraan-Korteweg \& Jerjen 2010, their fig.~3), with the northern component 
assumed to be the true stellar nucleus and the southern component a 
star-forming region. De Swardt et al. (2010) also estimated the total stellar 
mass of NGC59 to be $5\pm2 \times 10^8$ M$_{\odot}$, from their H band data. 
NGC59 is nearby, at a distance of only 4.4 Mpc and it is part of the 
Sculptor group of galaxies. 
The proximity and unusual stellar population in NGC59 make it 
an interesting early-type dwarf galaxy to study. More observations of 
ETGs in this low mass regime will help to constrain models of galaxy 
formation. 

The observations are described in section 2, then the data reductions are 
presented in section 3, including kinematics and emission-line strengths. 
Absorption lines are measured in section 4 and stellar population analysis 
is presented in section 5. Section 6 gives the conclusions.

\section[]{SALT Observations}

SALT is an 11 metre optical telescope that is optimised to be efficient in 
the blue region of the spectrum, thanks to its optical design and coatings. 
We used the Robert Stobie Spectrograph (RSS) on SALT to obtain medium 
resolution spectra across a broad wavelength range (Burgh et al. 2003, 
Buckley et al. 2008). For the extreme blue part of the spectrum we used 
the PG3000 grating, covering 3440 to 4220 \AA, with slit width 0.6arcsec, 
giving a resolution of 1.18 \AA\ (0.26 \AA/pix). For redder wavelengths we used 
the PG1300 grating, covering 3990 to 6080 \AA, with slit width 1.25arcsec, 
giving a resolution 
of 2.20 \AA\ (0.67 \AA/pix). For both the blue and red spectra the binning was 
a factor of 2 in the spectral direction and a factor of 4 in the spatial 
direction, giving 0.5068 arcsec per spatial bin along the slit. 
The data were observed between 24/05/2012 and 11/10/2012, including 
3 $\times$ 800 plus 1100 seconds in the blue and 3 $\times$ 900 seconds 
in the red, plus observations of three Lick standard stars in both blue 
and red spectral modes. All observations were performed in seeing 
conditions between 2 and 3 arcsec.
The slit position was along the minor axis of NGC59, at PA=37$\degr$ 
east of north (or 180 degrees around from there), for all spectra 
except for one of the PG3000 spectra which had PA=65$\degr$. These latter 
data were included to increase the signal-to-noise in the blue. 
Fig.~1 shows a red DSS image of the galaxy with the slit positions overlaid.
The inset is a zoom-in of the nucleus on our short 3 sec RSS acquisition 
image in white light. In poor seeing conditions a north-south elongation 
of the nucleus is just detectable, corresponding to the direction of the 
second nucleus apparent in fig.3 of de Swardt et al. (2010).

\begin{figure}
 \begin{center}
 \includegraphics[width=85mm, angle=0]{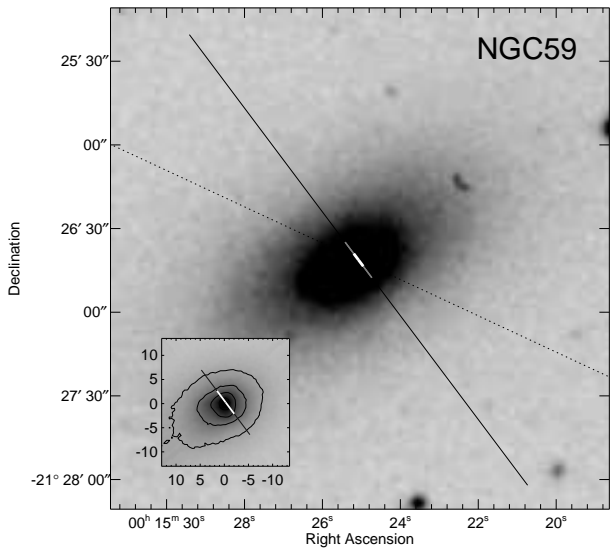} 
 \caption{ 
A 3$\times$3 arcmin photographic red DSS image of NGC59, north is up, 
east to the left.
The main RSS slit orientation PA=37$\degr$ is overlaid on the image with 
a solid line, and the extracted apertures indicated: the white part of 
the line is the Central region, and the grey sections of the line indicate 
the Lower (south west) and Upper (north east) apertures discussed in the 
text (Section 3.1). The PA=60$\degr$ slit is shown as the dotted line. 
The inset shows the nuclear area from the RSS acquisition image with 
clear filter, where 
some photometric asymmetry is evident. The inset tick marks are in arcsec.}
\end{center}
\end{figure}

The detector consists of three adjacent CCDs with two gaps. 
We chose wavelength ranges so as to minimize the loss of important 
regions of the spectra. 
Thus the blue spectrum covers higher order Balmer lines from H$\delta$ and 
above, plus the Ca H\&K features, whilst the red spectrum covers from H$\delta$ 
up to the NaD doublet. This then includes most of the standard Lick 
spectral indices, which are extensively used in stellar population analysis 
(Worthey et al. 1994, Proctor \& Sansom 2002, Johnsson et al. 2013), plus 
higher order Balmer lines, which are sensitive to the age of the most recent 
star formation (Worthey \& Ottaviani 1997).

\section[]{Data Reductions}

\subsection{Basic Reductions}

The preliminary reduced data obtained from the SALT pipeline 
(Crawford et al. 2010) were further 
reduced in IRAF. We divided by an illumination flat and wavelength calibrated 
the spectra, giving a typical uncertainty of $\pm 0.06$ \AA\ in the blue and 
$\pm 0.18$ \AA\ in the red. At the centres of these two spectral ranges this 
corresponds to velocity uncertainties of 4.7 and 10.7 kms$^{-1}$ respectively. 
The 2-d data were cosmic ray cleaned using {\it xzap} in IRAF and 
the spectra were background subtracted using a 2nd order polynomial. A relative 
flux calibration was carried out using flux standard star EG21, from 
observations taken as close in time as possible to the galaxy observations. 
This was done with a spline fit, order 6 applied to the flux standard data.
The galaxy spectra were then combined using {\it lscombine}, with 
additional cosmic ray cleaning, to give two cleaned, summed spectra, 
one in the blue and one in the red, with a total exposure of 3500 seconds 
in the blue spectrum and 2700 seconds in the red spectrum for the target 
galaxy.

A few cosmic ray features remained in the 2-d spectral images,
despite efforts to clean them automatically in the processing above.
Therefore,
we also carried out a final visual check of the spectral images and
patched out any remaining cosmic-ray effects that we could identify. 
This does not preclude the presence of faint cosmic rays, which would go 
unnoticed against the background of rapidly changing light across the 
spectral images for NGC59. 

Error arrays for the galaxy observations were generated initially from 
the counts in the original 2-d data files and errors were 
propagated at each step in the data reductions. The resultant random 
error arrays highlight where the noise is highest (e.g. due to strong 
sky lines, emission lines in the galaxy and increasingly towards the 
blue end of the PG3000 grating spectrum. These arrays were used in the 
subsequent analysis, including determination of signal-to-noise ratios.

Spectra were extracted for the three Lick calibration stars, which are also 
velocity standard stars, for the flux calibration stars and for the galaxy 
observations, in each of the blue and red spectral modes. 
The spectra for Lick stars and for the galaxy were flux calibrated using the 
standard star data. This is a relative flux calibration only, 
since absolute flux calibration is difficult with SALT and weather conditions 
cannot be assumed to be photometric.

Three spatial regions in NGC59 were measured:
\begin{enumerate}
  \item Central region covers r$_e$/8 either side of centre, which is 11 
spatial pixels across (5.6 arcsec), taking r$_e$ as 21.73 arcsec from RC3.
  \item Upper region covers 10 spatial pixels above the central region 
(5.1 arcsec).
  \item Lower region covers 10 spatial pixels below the central region 
(5.1 arcsec).
\end{enumerate}

The central spatial region corresponding to a radial extent of r$_e$/8 
($\sim$11 pixels in diameter) has a typical signal-to-noise ratio 
of S/N$\approx$47 in the blue spectrum and S/N$\approx$60 in the red spectrum, 
ignoring the blue regions (below $\sim 3800$ \AA, where the S/N drops rapidly).

\subsection{Kinematics and Emission Lines}

The Lick standard stars were first analysed in order to align them to a 
rest wavelength scale. The spectra were cross-correlated to determine 
their relative spectral shifts. The heliocentric velocity of one 
template star was found from the SIMBAD database and used to determine 
observed shifts for all the Lick 
standard stars. Once these shifts were known {\it dopcor} and {\it dispcor} 
in {\it iraf} were used to correct and resample the star spectra to a rest 
wavelength scale, with the same range and sampling as for the galaxy data. 
The aligned Lick star spectra could be used as templates for 
measuring galaxy kinematics, since the stars and galaxy were observed with 
the same setup. The penalized pixel fitting {\it pPXF} software (Cappellari 
and Emsellem 2004) was used to measure kinematics for NGC59, using the 
PG1300 data only. Emission lines, gap regions, residual sky lines and edges 
were masked out for these measurements. This gave a first velocity dispersion 
estimate of 27$\pm$11 kms$^{-1}$ and a measured recession velocity of 
332$\pm$9 kms$^{-1}$.

Since there were only three Lick standard stars observed, the template match
to NGC59 was not very accurate. To improve the accuracy of template stars 
to describe the absorption line spectrum in NGC59, template spectra of 
SSPs based on stars in the MILES spectral 
library from 2006 were used (Sarzi private communication).
To use the MILES spectra, the SALT spectra were Gaussian blurred to the same 
spectral resolution of 2.5 \AA\ FWHM (Falc\'{o}n-Barroso et al. 2011).
Running the {\it pPXF} software on these blurred SALT data for NGC59, 
with the MILES SSPs as templates led to the second velocity dispersion
estimate of 47$\pm$23 kms$^{-1}$.

Both of these estimates are similar to our previous estimate of 37 
$\pm$5 kms$^{-1}$, using data from the NTT with a large sample of 
Lick standard stars (Sansom \& Northeast 2008). For the rest of this 
paper we use this previous estimate of velocity dispersion, since it 
was obtained using more kinematic standard stars, giving a better 
template match to the galaxy at the resolution of those observations. 

In order to measure emission lines the two SALT gratings (PG3000 and PG1300) 
were first combined, from the same spatial region as far as possible, in 
the following way. Both the spectra were resampled onto the full spectral 
range (3440.6 to 6080.0 \AA) at the binning of the PG1300 spectra, then 
the overlap region (3990.0 to 4219.8 \AA) was normalised to the PG1300 
value, halved, added together with the shorter and longer wavelength data 
from single grating observations. This led to a spectrum that was well 
matched across the two grating observations. This spectrum was then 
blurred to the resolution of MILES, so that the full range of MILES SSP 
spectra could be used as templates during the emission line measurements. 
The spectrum of NGC59 was also binned to the MILES sampling of 0.9 \AA/pixel.
A best fit template spectrum (Best\_Template), combining several SSPs, 
was obtained by running {\it pPXF} with the emission lines masked out.
The Best\_Template was output as a spectrum, with Galactic 
extinction incorporated to match the observations, assuming 
Calzetti et al. (2000) reddening plus Galactic E(B-V)=0.019 mag, 
from NED\protect\footnotemark.

Emission lines in the spectrum of NGC\,59
were then accurately measured, making use of the Best\_Template 
and applying programs described in detail in \citet{SHOC,Sextans}.
Table~\ref{t:Intens} lists the measured relative
intensities of all detected emission lines relative to H$\beta$
(F($\lambda$)/F(H$\beta$)) and the ratios corrected for the extinction
(I($\lambda$)/I(H$\beta$)).
The EW of the H$\beta$ emission is also listed as is the 
derived extinction coefficient $C$(H$\beta$).
The latter is a sum of the internal extinction in NGC\,59
and foreground extinction in the Milky Way.
The measured $C$(H$\beta$) corresponds to a
V-band extinction of A$_{\rm V} = 0.59$~mag. Accounting 
for the Milky Way foreground extinction of 0.06~mag 
\citep{Schlegel98} this suggests non-negligible dust content of
A$_{\rm V} \sim 0.5$~mag extinction in the central HII region of NGC\,59.

Our emission line measurement programs determine the
location of the continuum, perform a robust noise estimation, and fit
separate lines by a single Gaussian superimposed on the
continuum-subtracted spectrum \citep{SHOC}. In this particular case
since a best-fit model was already created, the model was used as the continuum.
The quoted errors of single line intensities include
components summed up in quadrature \citep{SHOC}.
The total errors were propagated in the calculations and are included in the
uncertainties of the element abundances and all other derived parameters
presented here.
Note that our programs are designed to derive simultaneously 
both the extinction coefficient C(H$\beta$) and the absorption equivalent 
width EW$_{abs}$($\lambda$) for the hydrogen lines as described 
in \citet{Izotov94}. We can use this to check the consistency between
the model continuum and the emission lines. The result for absorption lines is
EW$_{abs}$($\lambda$)=0.00$\pm$0.13 \AA\ which means that the model that was
subtracted describes very well the Balmer absorption since no additional 
component is needed to explain the emission line ratios beyond extinction.
We experimented with slightly different models as well, and also by forcing 
EW$_{abs}$($\lambda$)=0, and the results for all emission line parameters stay 
the same within the uncertainties, which also stay the same.
We note that the uncertainty of 0.13
for the Balmer absorption is propagated through all the results, and, in fact, 
is very similar to the uncertainty of the Balmer line Lick indices derived 
below by very different and independent methods.
 
The HII region spectrum was interpreted by the technique of
plasma diagnostics and iterative calculations 
as described in detail in \citet{Ketal08}. The
results regarding temperatures, the number density $\rm n_e$ derived using 
the [OII] $\lambda$3726/$\lambda$3729 lines ratio, and
the total elemental abundances
for O, Ne and Cl, are given in Table~\ref{t:Chem}.
The electron temperature $T_{\rm e}$(OIII)
was calculated directly using the weak auroral line of oxygen
[O\,{\sc iii}] $\lambda$4363.

The relative emission line intensities given in Table.~1 are consistent 
with a star-forming origin, considering the location of [OIII]/H$\beta$ 
versus [OII]/[OIII] in the line ratio diagrams like those of Baldwin, 
Phillips and Terlevich (1981), their figure 2. This is consistent with the 
findings of previous authors.  Skillman, Cote \& Miller (2003) mapped and 
analysed the H$\alpha$ emission in NGC59. From H$\alpha$ emission and 
GALEX FUV fluxes, Karachentsev \& Kaisina 2013 estimated star formation 
rates of SFR$_{H\alpha}$=0.0123 M$_{\odot}$ yr$^{-1}$ and
SFR$_{FUV}$=0.0063 M$_{\odot}$ yr$^{-1}$ respectively, which agree quite 
well within their expected uncertainty of $\sim$50 per cent, confirming 
the star forming origin of the emission lines.
Bouchard et al. (2005) suggested that NGC59 be reclassified as dS0 Pec, 
due to the detection of neutral and ionised components near the centre. 
All the reddening corrected Balmer lines measured here (Table.~1), 
including the well determined higher order lines (H9 to H11), 
show ratios that are consistent with expectations for hydrogen 
recombination lines from a typical HII region (e.g. Skillman et al. 1994).

\begin{table}
\centering{
\caption{Emission line measurements relative to H$\beta$.
The relative intensities are measured from NGC\,59 after 
continuum model subtraction, using programs described in Kniazev et al. 
(2004, 2005). Note that the derived Balmer absorption EW(abs) $\approx 0$ 
indicates the absorption was correct in the continuum model (see text). 
Extinction is derived from a simultaneous iterative fit to the emission 
line intensities, absorption characteristics and temperature information using 
the direct method with the [O\,{\sc iii}] $\lambda$4363 Auroral line. 
All relevant errors have been propagated to the uncertainties including 
those from the continuum fitting.}
\label{t:Intens}
\begin{tabular}{lcc} \hline
\rule{0pt}{10pt}
$\lambda_{0}$(\AA) Ion        & F($\lambda$)/F(H$\beta$)&I($\lambda$)/I(H$\beta$) \\
                              & (measured) & (corrected) \\ 
\hline
3727\ [O\ {\sc ii}]\          & 1.8667$\pm$0.0857 & 2.2846$\pm$0.1089 \\
3771\ H11\                    & 0.0140$\pm$0.0040 & 0.0170$\pm$0.0070 \\
3798\ H10\                    & 0.0214$\pm$0.0034 & 0.0259$\pm$0.0065 \\
3835\ H9\                     & 0.0513$\pm$0.0032 & 0.0614$\pm$0.0061 \\
3868\ [Ne\ {\sc iii}]\        & 0.4282$\pm$0.0095 & 0.5094$\pm$0.0125 \\
3889\ He\ {\sc i}\ +\ H8\     & 0.1847$\pm$0.0049 & 0.2189$\pm$0.0079 \\
3967\ [Ne\ {\sc iii}]\ +\ H7\ & 0.2315$\pm$0.0103 & 0.2702$\pm$0.0133 \\
4026\ He\ {\sc i}\            & 0.0146$\pm$0.0011 & 0.0169$\pm$0.0012 \\
4068\ [S\ {\sc ii}]\          & 0.0227$\pm$0.0026 & 0.0260$\pm$0.0030 \\
4076\ [S\ {\sc ii}]\          & 0.0041$\pm$0.0009 & 0.0047$\pm$0.0011 \\
4101\ H$\delta$\              & 0.2550$\pm$0.0058 & 0.2902$\pm$0.0083 \\
4340\ H$\gamma$\              & 0.4220$\pm$0.0094 & 0.4600$\pm$0.0112 \\
4363\ [O\ {\sc iii}]\         & 0.0492$\pm$0.0017 & 0.0535$\pm$0.0019 \\
4471\ He\ {\sc i}\            & 0.0297$\pm$0.0017 & 0.0317$\pm$0.0018 \\
4861\ H$\beta$\               & 1.0000$\pm$0.0294 & 1.0000$\pm$0.0297 \\
4922\ He\ {\sc i}\            & 0.0081$\pm$0.0011 & 0.0081$\pm$0.0011 \\
4959\ [O\ {\sc iii}]\         & 1.5665$\pm$0.0335 & 1.5432$\pm$0.0332 \\
5007\ [O\ {\sc iii}]\         & 4.7950$\pm$0.1026 & 4.6896$\pm$0.1009 \\
5518\ [Cl\ {\sc iii}]\        & 0.0046$\pm$0.0011 & 0.0042$\pm$0.0010 \\
5538\ [Cl\ {\sc iii}]\        & 0.0028$\pm$0.0008 & 0.0025$\pm$0.0007 \\
5876\ He\ {\sc i}\            & 0.1212$\pm$0.0029 & 0.1054$\pm$0.0027 \\
  & & \\
C(H$\beta$)\ dex          & \multicolumn{2}{c}{0.28$\pm$0.04} \\  
E(B-V)\ mag               & \multicolumn{2}{c}{0.19$\pm$0.03} \\
A$_V$\ mag                & \multicolumn{2}{c}{0.59$\pm$0.08} \\
EW(abs)\ \AA\             & \multicolumn{2}{c}{0.00$\pm$0.13} \\   
EW(H$\beta$)\ \AA\        & \multicolumn{2}{c}{  44$\pm$ 1}   \\   
\hline
\end{tabular}
 }
\end{table}

\begin{table}
\centering{
\caption{Elemental abundances in NGC\,59.}
\label{t:Chem}
\begin{tabular}{lc} \hline
$T_{\rm e}$(OIII)(K)\                & 12,139$\pm$173 ~~   \\
$T_{\rm e}$(OII)(K)\                 & 11,952$\pm$149 ~~   \\
$N_{\rm e}$(OII)(cm$^{-3}$)\         &  48$\pm$33 ~~       \\
& \\
O$^{+}$/H$^{+}$($\times$10$^5$)\     & 4.477$\pm$0.297~~   \\
O$^{++}$/H$^{+}$($\times$10$^5$)\    & 9.161$\pm$0.418~~   \\
O/H($\times$10$^5$)\                 & 13.640$\pm$0.513~~  \\
12+log(O/H)\                         & ~8.13$\pm$0.03~~    \\
& \\ 
Ne$^{++}$/H$^{+}$($\times$10$^5$)\   & 2.678$\pm$0.149~~   \\
ICF(Ne)\                             & 1.141               \\
Ne/H($\times$10$^5$)\                & 3.06$\pm$0.17~~     \\ 
12+log(Ne/H)\                        & 7.49$\pm$0.02~~     \\ 
log(Ne/O)\                           & $-$0.65$\pm$0.03~~  \\
& \\ 
Cl$^{++}$/H$^{+}$($\times$10$^7$)\   & 0.291$\pm$0.056~~   \\
ICF(Cl)\                             & 1.286               \\
Cl/H($\times$10$^7$)\                & 0.37$\pm$0.07~~     \\ 
12+log(Cl/H)\                        & 4.57$\pm$0.08~~     \\ 
log(Cl/O)\                           & $-$3.56$\pm$0.08~~  \\
\hline
\end{tabular}
 }
\end{table}

Abundances in the gas were estimated from the observed 
emission lines, using the software of Kniazev, with the 
direct method because the [OIII] line at 4363 \AA\ is detected. 
The oxygen abundance 12+log(O/H) = 8.13$\pm$0.03 dex 
(i.e.$\sim$1/5 Solar, e.g. compared to 
Grevesse \& Sauval 1998) that we find for NGC\,59 is 
reasonably consistent with the abundance 
8.29$\pm$0.08 dex published by \citet{Setal08}.
Our data also allow us, for the first time, to determine 
abundances of Ne and Cl in NGC\,59, which are also 
approximately 1/5 Solar to within about 20\%. 
It is known that HII region abundances mainly provide 
information about $\alpha$-process elements, which are 
produced predominantly in short-lived massive stars.
The relative abundances of log(Ne/O) and log(Cl/O) in Table.~1
are consistent with values for a large sample of 
HII regions in blue compact galaxies \citet{IT99}
and a large sample of HII galaxies \citep{ISMGT06} from
the Sloan Digital Sky Survey DR3 data \citep{DR3}.

\begin{figure*}
 \centering
 \begin{minipage}{150mm}
 \begin{center}
 \includegraphics[width=110mm, angle=-90]{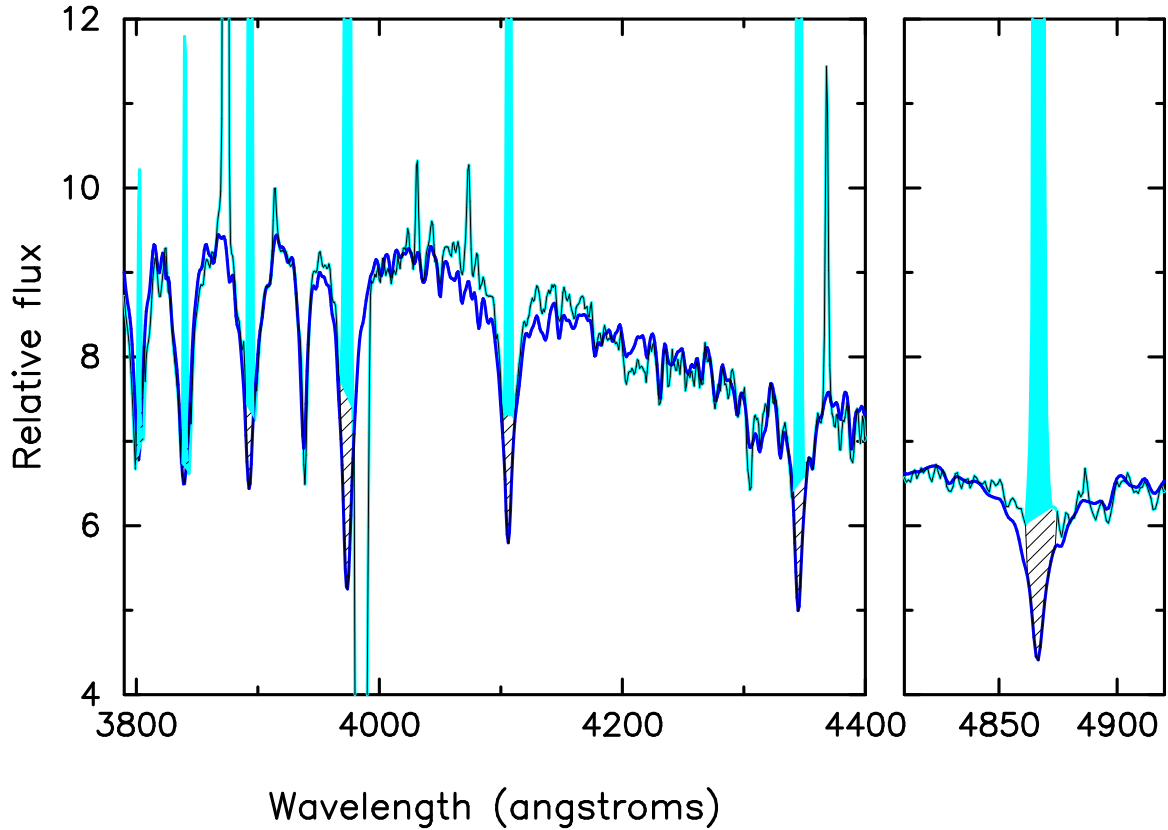} 
 \caption{ 
Balmer line regions in the central spectrum in NGC59, illustrating the replacement of 
the cores of Balmer lines, where strong emission lines originally dominated those data. 
The thick cyan lines are the original spectral data, the thick dark blue lines are the
best fitting MILES template data and the thin black lines are the Balmer line core
corrected spectral data. Other emission lines are subsequently removed by interpolation 
across them, to produce the lower spectrum shown in Fig.~3.}
\end{center}
\end{minipage}
\end{figure*}

\subsection{Removing Emission Lines}

Subtracting the emission lines proved difficult to do accurately, 
since emission lines near the centre of NGC59 are quite 
strong and their spectral shape in the SALT data did not appear to follow 
a Gaussian accurately enough to subtract them well using {\it gandalf}
(Sarzi et al. 2006).
An alternative solution was used to remove the estimated contributions 
from emission lines. 
For all hydrogen Balmer lines in the spectral range, core regions 
(width $\sim 10$ \AA) affected by emission were 
replaced by data from Best\_Template, as our best estimate of the shape 
and strength of the hydrogen absorption line cores. This is 
illustrated in Fig.~2, which shows a closeup of the Balmer line regions, 
illustrating the original data (thick cyan line), the best fit template 
(thick dark blue line) and the data with Balmer cores replaced (thin black line).
For other emission lines, simple interpolation across 
the lines was applied and these regions flagged for subsequent information.
Fig.~3 shows the original central spectum (top plot), then the final central 
spectrum with the hydrogen line emission replaced and other 
emission lines removed (lower plot). Although we are replacing part of the 
hydrogen lines by template data, rather than real data for NGC59, 
this produces more realistic hydrogen absorption line shapes than 
subtracting the emission lines using {\it gandalf}. This replacement 
process was thought to be the best compromise, given that the core 
absorption line data is lost due to the strong emission lines. 
H7 at 3970 \AA\ is lost in one of the gaps, but many other Balmer lines 
are clearly visible. The consistency of measured higher order Balmer 
line emissions (shown in Table.~1) helps to support the validity of our 
approach. Higher order Balmer absorption lines are less affected by 
emission line contamination.

\footnotetext{NED is the Nasa Extragalactic Database at http://ned.ipac.caltech.edu}

At short wavelengths there is a hint of UV upturn in the spectrum 
(Kaviraj et al. 2007), however near the blue edge of the data, below 
$\sim$3500 \AA\ the flux calibration becomes unreliable.

\begin{figure}
 \begin{center}
 \includegraphics[width=75mm, angle=90]{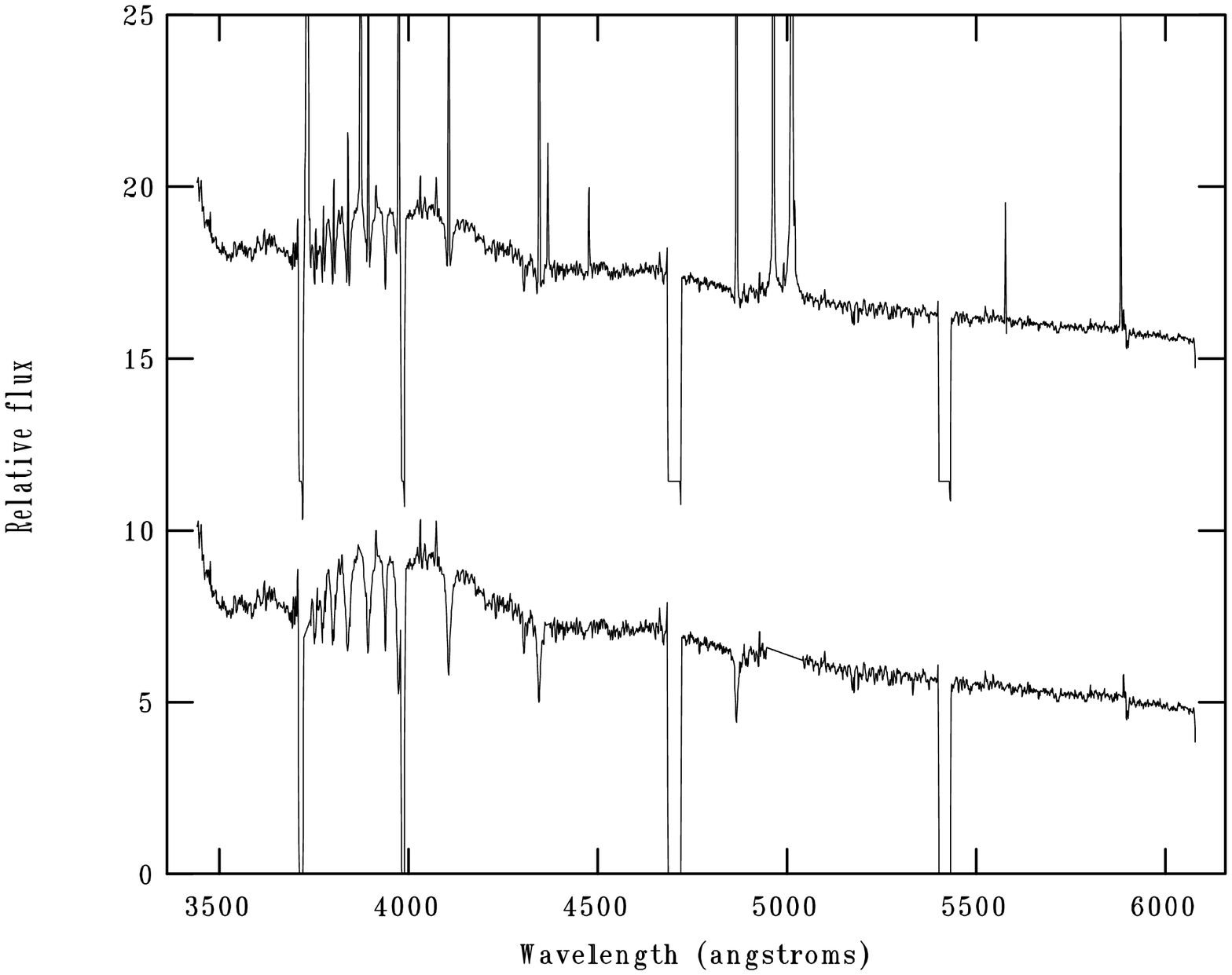} 
 \caption{ 
Central spectrum in NGC59. This plot shows the processing to remove 
emission lines from the spectrum. Data were taken with the SALT telescope 
in two grating settings, combined as described in the text. The {\bf top plot} 
is offset vertically and shows the combined data before attempting to 
remove any emission line 
features. This shows four gap regions, two associated with each grating 
observation, centred around 3716 \AA\ and 3985 \AA\ in the blue, and around 
4703 \AA\ and 5417 \AA\ in the red. Emission lines are strong and 
the hydrogen Balmer series 
shows both emission lines and broader absorption line components. 
Residual sky line features are apparent at 5577 \AA\ and residual 
contamination due to strong cosmic rays affects the NaD doublet such 
that we do not use it in our subsequent analysis in section 5. 
The {\bf lower plot} shows the same spectrum with the Balmer emission 
line regions replaced by data from the best fit MILES SSP template and
several strong emission lines removed by interpolating across those regions.}
\end{center}
\end{figure}

\section[]{Absorption Line Measurements}

Standard Lick absorption lines were measured using the {\it Lector} software 
available from http://miles.iac.es/pages/software.php. Perturbations 
applying the error arrays were used to generate 500 Monte Carlo realisations 
of the data, from which random errors were estimated for the line-strength 
meaurements. Line-strength errors due to uncertainties in kinematics 
(velocity and velocity dispersion) were evaluated by varying the assumed 
kinematic within their estimated errors and re-measuring the line strengths.
These uncertainties were found to be significantly less than the random errors.
Uncertainties in the Balmer line cores were estimated by trying fits using 
different parts of the spectrum, but still including all the Balmer lines,
including removing the continuum regions above H$\beta$ and below the Balmer 
limit. This led to estimated Balmer line errors from our core H line 
replacement procedure.
All these errors were added together in quadrature to give total estimated 
errors for each index.
Two Lick features were 
lost due to partially falling within gap regions (C$_2$4668 and Fe5406). 
Another Lick feature was lost due to falling too close to strong emission 
lines (Fe5015). The data and error estimates are given in Table.~3, for
the three regions measured in NGC59.

Fig.~4 shows a plot of these three regions. Evidence of young stars is 
present in all three spectra, from the strength of the Balmer absorption 
lines. From the cleaned spectra in Figs.~3 and 4 it is already possible 
to see that this is a low metallicity galaxy, from the weakness of the 
metal sensitive lines around Mgb.
One possibility to consider for NGC59 is that nebular continuum emission
may decrease its apparent absorption line strengths. Reines et al. (2010)
studied relative contributions to continuum emission from stars and gas 
in young massive star clusters. For example, their figure 8 shows that 
if all the stars had formed only 5 Myr ago then the nebular continuum is 
only ~5 per cent of the total continuum. The results presented here 
(e.g. Fig.~3), including strong Balmer absorption lines in NGC59, show 
that stars older than 5 Myr dominate the light. This is also supported by 
full spectrum fitting described below in Section 5.4. Therefore the effect 
of nebular continuum emission is unlikely to cause a significant bias 
for the Lick absorption line measurements in NGC59.

\begin{figure}
 \begin{center}
 \includegraphics[width=75mm, angle=90]{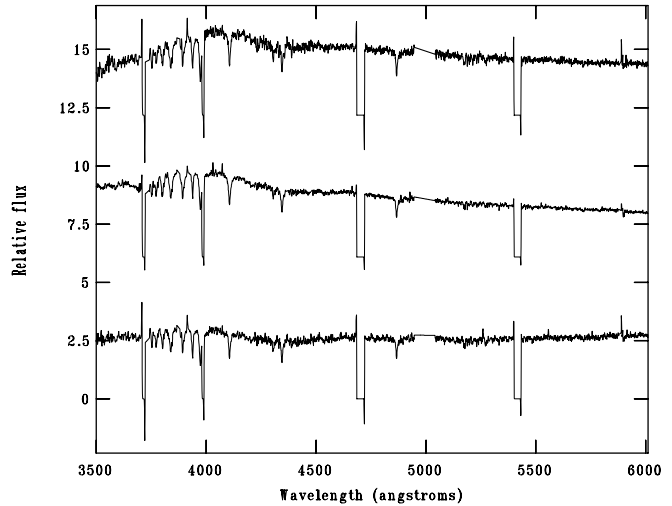} 
 \caption{ From bottom to top, this plot shows cleaned spectra of the 
lower, central and upper regions in NGC59, with vertical offsets to 
separate the plots. The central region 
is the same as in the lowest plot in Fig.~3. All three spectra have had 
the cores of their hydrogen Balmer lines replaced by the best fitting SSP 
template from the MILES SSP database, in order to remove the emission line 
contamination from those lines. Other strong emission lines have been removed 
by interpolating across them. 
}
\end{center}
\end{figure}

The upper, off-centre region extends beyond the two peaks of near 
infrared emission mapped by de Swardt et al. (2010), which they assumed 
to be associated with the centre of the galaxy and a star-forming 
peak 2.3 arcsec to the south. Therefore it is interesting to see if 
different stellar population characteristics are uncovered in these 
regions beyond the central star-forming area, perhaps better 
characterising the bulk of stars in NGC59.

Table.~3 shows how the spectra change between central and 
off-centre regions, with hydrogen absorption lines becoming generally 
weaker in the lower region. There is an anomalously weak 
measurement of Fe5270 in the lower region, which is 
investigated below. These data are used to characterise 
the stellar populations in the next section.

\section[]{Stellar Population Analysis}

\subsection{Lick System - MILES SSPs}

The indices from Table.~3, at the standard Lick resolutions 
(Worthey \& Ottaviani 1997), were used to estimate luminosity weighted 
average SSP age, metallicity and [$\alpha$/Fe] across NGC59. This uses 
software written by R. Proctor, based on the SSP results of Korn et al. 
2005, plus references therein. 
CN1 was excluded from these fits since it is too strongly affected by 
H$\delta$ in its blue side band. Results are given in Table.~4.

The fit for the lower region led to an old age, 
qualitatively inconsistent with the presence of emission lines 
in this spectrum, suggesting some more recent or ongoing star formation. 
The initial fit was quite poor due to an anomalously low value 
of Fe5270, which when removed led to a significant improvement 
in chi-squared at similar parameters. 
On further inspection, a small peak in the lower spectrum, 
in the central band of Fe5270 adversely affects this measurement and 
appears to correspond to a possible weak cosmic ray event in one of 
the original images. We note that removing a remaining outlier, 
the Mg1 index, leads to a younger estimated age (Log(Age)=0.8 dex) 
while the metallicity remains low and the abundance ratio sub-solar. 
This gives an indication of the parameter uncertainties due to the 
inclusion or exclusion of specific indices. The fit with Fe5270 
removed is shown in Table.~4.
For this low metallicity ([Z/H]$\sim -1.6$) the SSP 
line-strengths change little with age, for ages greater than 8 Gyrs 
(see Fig.~5) and the Balmer line-strengths start to vary 
non-monotonically. Therefore the best fitting SSP age of this 
lower region is difficult to constrain beyond 
saying that it is older than $\sim$ 8 Gyrs. The uncertainty 
in age estimate given for this lower region in Table.~4 is an 
underestimate since we can only give this lower limit to the age.
Therefore we do find evidence of older stars ($>\sim$ 8Gyr) in NGC59, 
beyond the central, strongest emission-line region.
Jerjen, Binggeli \& Freeman (2000) studied its optical
brightness profile and colours and found a Sersic index of 
n$=0.63$ in B, with B-R colour becoming redder further out, 
indicative of older stars.

The centre region was not well fit by a single SSP ($\chi_r^2=2.63$).
Outlying indices were removed to see what effect this 
would have on the fit and derived parameters. Removing 2 indices 
(Mg1 and Fe5270) brought the $\chi_r^2$ down to 1.4, but 
did not significantly alter the derived parameter values. Therefore 
we show the fit including those three indices in Table.~4.
In Section 5.3 below we make a preliminary search to see if two SSPs
will better describe this central region of NGC59.
The metallicity of the centre region is similar to that in the 
lower region. 
The low [Fe/H] abundance found in the centre of NGC59 is also supported 
by the Rose indices measured using {\it Lector} in this region. 
For NGC59 we measured 
H$\delta$/FeI$\lambda$4045=0.654 and FeI$\lambda$4045/FeI$\lambda$4063=1.012, 
which implies [Fe/H]$\sim<-1.5$ (Rose et al. 1994, their fig.~10).
The centre regions has a fitted SSP age of $\sim 5$Gyr.

\begin{landscape}

\begin{table}
\footnotesize
 \centering
  \caption{
NGC59 line strengths measured at {\bf MILES resolution} 
(FWHM=2.5 \AA), for standard Lick band definitions. Standard deviations (SD) 
are 1 sigma random errors derived from 500 realisations of the spectra, 
perturbed by the error array. Dsig errors are derived by accounting for
the uncertainty in velocity dispersion estimated from the current data. 
Similarly, Dvel errors are derived by 
measuring the changes in indices considering the uncertainty in
recession velocity. H errors are estimated uncertainties in the Balmer 
line core replacment. Total errors add all these error sources together,
in quadrature. The errors are dominated by the random errors and by Balmer 
line core uncertainties. The mask values are '1' for good indices, included 
in later fits, and '0' for indices with problems, excluded from later fits. 
These latter indices include: C$_2$4668 and Fe5406, which fall into the 
gap regions between CCDs; Fe5015, which overlaps regions containing strong
emission lines. The line strengths for these three indices are not shown 
here. This mask array applies to all three spatial regions, unless otherwise
stated in the text.
Also tabulated are results for the same three regions, at the 
{\bf Lick standard resolution} (from Worthey \& Ottaviani 1997) at 
FWHM $\sim$8.4 - 10.9 \AA.
}
  \setlength{\tabcolsep}{1.0mm}
  \begin{tabular}{@{\hspace{0.1mm}}
l@{\hspace{3.0mm}}
ccccccccccccccccccc@{\hspace{0.1mm}}}
  \hline
Name:        &H$\delta_A$  &H$\delta_F$  &CN1  &CN2  &Ca4227  &G4300  &H$\gamma_A$  &H$\gamma_F$  &Fe4383  &Ca4455  &Fe4531  &H$\beta$ &Mg1  &Mg2  &Mgb  &Fe5270  &Fe5335  &Fe5709  &Fe5782 \\ 
\hline
\multicolumn{20}{l}{\bf Central (r$_e$/8) region. MILES resolution.} \\  
Value:       & 4.856 & 3.977 &-0.098 &-0.051 & 0.176 & 0.475 & 3.254 & 3.132 & 0.578 & 0.395 & 1.286 & 3.103 & 0.020 & 0.041 & 0.962 & 0.657 & 0.688 & 0.473 & 0.287 \\
SD err:    & 0.163 & 0.116 & 0.005 & 0.006 & 0.092 & 0.168 & 0.176 & 0.124 & 0.261 & 0.121 & 0.180 & 0.170 & 0.003 & 0.003 & 0.126 & 0.144 & 0.153 & 0.103 & 0.102 \\
Dsig err:  & 0.003 &-0.003 & 0.000 & 0.000 &-0.001 & 0.000 &-0.002 &-0.014 &-0.052 & 0.016 &-0.006 & 0.003 &-0.001 &-0.001 &-0.007 &-0.003 &-0.008 &-0.001 & 0.005 \\
Dvel err:  & 0.000 & 0.003 & 0.000 &-0.001 & 0.005 &-0.026 & 0.001 & 0.002 &-0.003 &-0.008 &-0.005 &-0.000 & 0.000 &-0.000 & 0.009 &-0.007 &-0.016 & 0.003 &-0.004 \\
H err:     & 0.638 & 0.643 &       &       &       &       & 0.428 & 0.428 &       &       &       & 0.499 &       &       &       &       &       &       &       \\
Total err: & 0.658 & 0.653 & 0.005 & 0.006 & 0.092 & 0.170 & 0.463 & 0.446 & 0.266 & 0.122 & 0.180 & 0.527 & 0.003 & 0.003 & 0.126 & 0.145 & 0.154 & 0.103 & 0.102 \\
\hline
Mask:  & 1     & 1     & 0     & 1     & 1     & 1     & 1     & 1     & 1     & 1     & 1     & 1     & 1     & 1     & 1     & 1     & 1     & 1     & 1 \\
\hline
\multicolumn{20}{l}{\bf Upper region. MILES resolution.} \\
 Value:       & 4.636 & 4.013 &-0.124 &-0.081 & 0.948 & 1.469 & 2.601 & 3.458 & 2.468 & 0.226 & 1.445 & 3.488 & 0.025 & 0.074 & 1.227 & 1.890 & 1.282 & 0.483 & 0.391 \\ 
Total err: & 0.761 & 0.735 & 0.010 & 0.012 & 0.185 & 0.357 & 0.463 & 0.553 & 0.507 & 0.233 & 0.378 & 0.484 & 0.005 & 0.006 & 0.241 & 0.274 & 0.297 & 0.187 & 0.193 \\
\hline
\multicolumn{20}{l}{\bf Lower region. MILES resolution.} \\
 Value:       & 4.143 & 3.527 &-0.077 &-0.026 & 0.434 & 1.162 & 2.344 & 3.167 & 1.850 &-0.212 & 1.439 & 2.436 & 0.032 & 0.063 & 0.919 & 0.457 & 0.732 & 0.638 & 0.177 \\
Total err: & 0.719 & 0.684 & 0.010 & 0.012 & 0.173 & 0.329 & 0.375 & 0.656 & 0.441 & 0.232 & 0.346 & 0.408 & 0.005 & 0.005 & 0.196 & 0.228 & 0.251 & 0.141 & 0.141 \\
\hline
\multicolumn{20}{l}{\bf Central (r$_e$/8) region. Lick Standard resolution.} \\  
Value:       & 4.815 & 3.644 &-0.097 &-0.065 & 0.084 & 0.290 & 3.194 & 2.899 & 0.468 & 0.267 & 1.088 & 2.996 & 0.020 & 0.041 & 0.830 & 0.558 & 0.568 & 0.435 & 0.274 \\ 
Total err: & 0.658 & 0.653 & 0.005 & 0.006 & 0.095 & 0.171 & 0.463 & 0.443 & 0.277 & 0.124 & 0.200 & 0.526 & 0.003 & 0.003 & 0.127 & 0.139 & 0.158 & 0.103 & 0.100 \\
\hline
\multicolumn{20}{l}{\bf Upper region. Lick Standard resolution.} \\
 Value:       & 4.670 & 3.636 &-0.127 &-0.099 & 0.667 & 1.128 & 2.617 & 3.215 & 2.298 & 0.056 & 1.369 & 3.409 & 0.025 & 0.073 & 1.029 & 1.747 & 1.057 & 0.484 & 0.297 \\
Total err: & 0.713 & 0.736 & 0.010 & 0.012 & 0.191 & 0.350 & 0.455 & 0.549 & 0.490 & 0.236 & 0.396 & 0.481 & 0.006 & 0.006 & 0.242 & 0.241 & 0.301 & 0.201 & 0.179 \\ 
\hline
\multicolumn{20}{l}{\bf Lower region. Lick Standard resolution.} \\
 Value:       & 4.163 & 3.210 &-0.079 &-0.044 & 0.261 & 0.926 & 2.275 & 2.911 & 1.669 &-0.168 & 1.246 & 2.493 & 0.031 & 0.062 & 0.871 & 0.371 & 0.610 & 0.516 & 0.187 \\
Total err: & 0.723 & 0.685 & 0.010 & 0.012 & 0.178 & 0.333 & 0.360 & 0.650 & 0.479 & 0.228 & 0.357 & 0.411 & 0.004 & 0.005 & 0.186 & 0.219 & 0.235 & 0.139 & 0.137 \\
\hline
\end{tabular}
\end{table}

\end{landscape}

\begin{table*}
 \centering
 \begin{minipage}{140mm}
  \caption{SSP fits from MILES SSP templates at the Lick standard resolution. 
Note that [Fe/H] is estimated from [Z/H] and [$\alpha$/Fe].}
  \begin{tabular}{@{}llcccccl@{}}
  \hline
 Region        & Log(Age) &  [Fe/H]  &  [$\alpha$/Fe]  &  [Z/H]  & No.  & Chi-sq  & Comment \\
               & (Gyr)    &          &          &          & ind. & reduced &         \\
               & ($\pm$)  & ($\pm$)  &  ($\pm$) &  ($\pm$) &      &         &         \\
 \hline
 Upper         &  0.550   &  -0.767  &  -0.300  &  -1.050  &  18  &  23.43  &  Emission line region  \\
               &  0.118   &   0.135  &   0.054  &   0.120  &      &         &         \\
 Centre        &  0.700   &  -1.412  &  -0.120  &  -1.525  &  18  &  44.63  &  Stronger emission line region \\
               &  0.047   &   0.211  &   0.177  &   0.089  &      &         &         \\
 Lower         &  1.175   &  -1.405  &  -0.180  &  -1.575  &  17  &  26.64  &  Emission line region; Fe5270 excluded \\
               &  0.082   &   0.236  &   0.192  &   0.122  &      &         &         \\
\hline
\end{tabular}
\end{minipage}
\end{table*}

\begin{figure*}
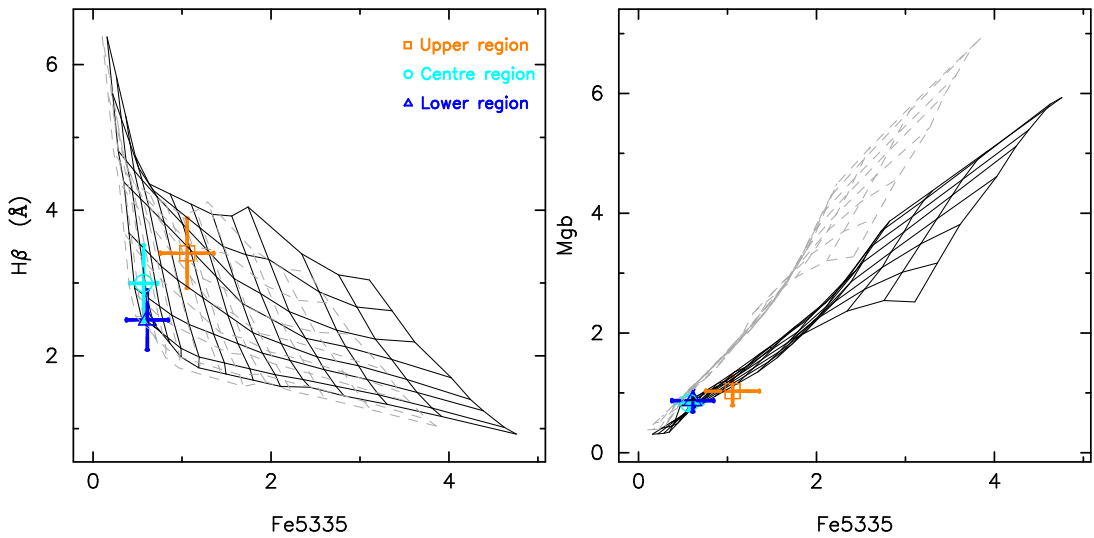

 \centering
 \begin{minipage}{160mm}
 \begin{center}
 \includegraphics[width=70mm, angle=-90]{pgplot_Fe5335vHb_symbols.ps}
 \includegraphics[width=70mm, angle=-90]{pgplot_Fe5335vMgb_symbols.ps}
 \caption{ 
Index results for NGC59, plotted against grids from MILES SSPs.
The black grid is for [$\alpha$/Fe]=0, while the grey grid is for 
[$\alpha$/Fe]=+0.3.
In the {\bf left plot}, the age indicator H$\beta$ is plotted, 
with age increasing downwards on the grid,for 
Log(Age(Gyr))=0.0,0.15,0.3,0.475,0.7,0.9,1.075,1.175.
The metallicity indicator Fe5335 is plotted, with metallicity increasing
from left to right, from [Z/H]$=-2.25$ to $+0.80$, in steps of 0.25 dex. 
Solar metallicity is the fourth line from the right.
This plot of age versus metallicity sensitive features, shows
the low metallicity and relatively young age in NGC59, with the upper 
region (orange) being the youngest and the lower region being the 
oldest (dark blue). The centre of NGC59 (cyan) is intermediate in age 
between the two outer regions. 
The {\bf right plot}, of Fe5335 versus Mgb, highlights the un-enhanced 
[$\alpha$/Fe] ratios in NGC59.
}
\end{center}
\end{minipage}
\end{figure*}

The upper region is  adequately fit by a single SSP 
($\chi_r^2=1.4$), which is relatively young (3.5 Gyr).
The upper region is fit by the highest overall metallicity 
([Z/H]$\sim -1.05$), which is still low amongst ETGs. 
The best fit [$\alpha$/Fe] ratio is sub-solar in all 
regions, similar to what we found previously with the NTT data 
(Sansom \& Northeast 2008), where only one spectrum was analysed.

In general the ages and abundance ratios are not very well 
constrained by these data. However, the very low metallicity 
in NGC59 ([Z/H] $<$ 1/10th of solar) is a more robust result. 
These results are illustrated in the index-index plots of specific 
features shown in Fig.~5, where the locations and uncertainties of 
the three regions in NGC59 are plotted against grids of MILES SSP 
predictions.

\subsection{Lick System - EZ\_Ages}
 
The above results were tested using an alternative stellar library 
and fitting software called EZ\_Ages (Graves \& Schiavon 2008), 
downloaded from http://astro.berkeley.edu/$\sim$graves/ez\_ages.html. 
This code uses SSPs based on the Jones and INDO-US stellar libraries, 
as described in Schiavon (2007) and fits to Lick standard indices.
Briefly, at solar abundance pattern, the ages searched are:
1.2, 1.5, 2.5, 2.8, 3.5, 5.0, 7.0, 10.0, and 14.1 Gyrs, and the 
[Fe/H] values searched are: $-1.3$, $-0.7$, $-0.4$, 0.0 and $+0.2$ dex. 
EZ\_Ages could not be used to fit indices for NGC59 in the central 
and lower regions because this galaxy has a very low metallicity, 
falling below the lowest predictions, at [Fe/H]$=-1.3$, in EZ\_Ages. 
This is illustrated in the output index-index plots, an example of 
which is shown in Fig.~6. Lines of constant [Fe/H] (solid lines) are 
shown, with [Fe/H] increasing from left to right; plus lines of 
constant age (dotted lines) have ages increasing from top to bottom 
in these grid plots. Therefore this figure indicates that the ages 
in NGC59 are consistent with $\sim$2 to 6 Gyrs, based on H$\beta$ 
as the main age indicator and slightly extrapolating to the left of 
the plotted grid (i.e. to lower [Fe/H]) for the centre and lower 
regions. For the upper region, which could be fitted with the EZ\_Ages 
software, the following measurements were found: 
Age=$2.12_{-0.45}^{+1.09}$ Gyr; [Fe/H]=$-0.80_{-0.24}^{+0.26}$ dex; 
[Mg/Fe]=$-0.04_{-0.09}^{+0.12}$ dex.

Results for ages and [Fe/H] from the MILES SSP fits are qualitatively 
consistent with those from EZ\_Ages. For the upper region the results 
from EZ\_Ages are also quantitatively consistent within the estimated 
errors given in Table.~4. One exception is that the best fit age from 
MILES fits to the lower region is very old. However, recall that 
we can only place a limit on this age as approximately $>$8 Gyrs, 
as discussed in Section 5.1 above. The good agreement between results 
from MILES SSP fits and EZ\_Ages fits indicates that our SSP fitting 
results are not biased by the models used to fit the Lick indices.

\begin{figure*}
 \centering
 \begin{minipage}{150mm}
 \begin{center}
 \includegraphics[width=120mm, angle=0]{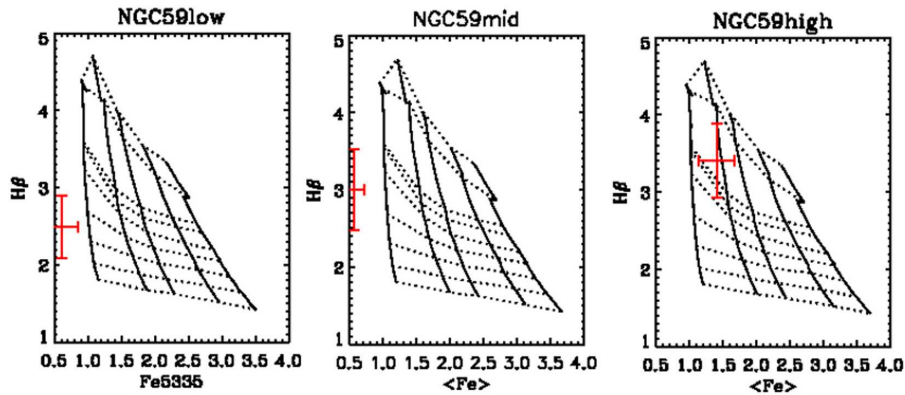} 
 \caption{ 
Grid plot showing results for NGC59 from EZ\_Ages fitting routine,
for the three spatial regions across NGC59. The Lick resolution data 
and errors are shown by the red plus sign. The upper region in NGC59 
(right plot) can be fit, with a metallicity lying within the range 
allowed by EZ\_Ages. The lower and centre regions in NGC59 (left and 
middle plots) cannot be fit, because of the low metal line 
strengths, indicating metallicities below those allowed for in 
EZ\_Ages (i.e. [Fe/H]$<-1.3$). The age indicator (H$\beta$) is 
plotted on the vertical axes, whilst metallicity indicators are 
plotted on the horizontal axes ($<$Fe$>=$Fe5270$+$Fe5335 or Fe5335).
}
\end{center}
\end{minipage}
\end{figure*}

\subsection{New MILES database - Two burst models}

We know that galaxies are not SSPs in reality, and that NGC59 shows 
evidence of very recent star formation in the form of strong emission 
lines. Here we generate some predictions for two-burst SFHs, using the 
webtools and the latest version of SSPs from the MILES website 
(Vazdekis et al. 2010), at the MILES resolution (FWHM=2.5 \AA). 
Composite spectra and Lick indices can be generated and downloaded.

We attempt a search of parameter space to see if the current data 
for NGC59 are consistent with an old ($\sim$14 Gyr), metal-poor 
([Z/H]$\sim -1.7$) burst plus a younger burst, whose age, metallicity and 
mass contribution we vary on a grid of models. Younger burst ages are 
from 0.063 to 1.0 Gyrs, with possible metallicities of 
[Z/H] = $-1.7$, $-1.3$, $-0.7$, $-0.4$, $0.0$ and mass contribution varied
from 50 per cent to 0.5 per cent. The rest of the mass is in the old SSP. 
These combinations were tested against the data for the centre region 
in NGC59. The best fits are restricted by the fact that no interpolation 
of SSP models was available from the MILES web site. 
A similar fit to that shown in Table.~4 could be achieved, but only with
the lowest metallicity ([Z/H]$=-1.7$), and a small mass contribution 
($\sim$ 1.5 per cent) from a young population (0.07 Gyr). This is equivalent
to a frosting of young stars on an otherwise old, metal-poor stellar 
population. The fit became worse if a higher metallicity ([Z/H]$=-0.7$) 
was assumed for the old or young population. Therefore only fits with low 
metallicity were possible, supporting the results from single SSP 
fitting that NGC59 is a very low metallicity galaxy.

Attempting to fit more complex SFH, with intermediate age populations 
led to poorer fits, since intermediate age populations cannot reproduce 
weak enough metal sensitive lines, at a given Balmer line-strengths.

\subsection{STARLIGHT full spectrum fits}

Finally, as an additional check on the analysis, both for Balmer line 
processing and to check model fits, full spectrum fitting was attempted 
using STARLIGHT (Cid Fernandes et al. 2005). This was run with 
Bruzual \& Charlot (2003) SSPs, for all the three apertures and with 
and without the cores of the strongest Balmer lines for each. Whether 
the Balmer lines were masked out or not, the results are very 
similar illustrating that our main results are not biased by the 
Balmer emission line replacement process described in Section 3.3.

Also the STARLIGHT fits are broadly consistent with the results from 
fitting Lick indices. The mean light-weighted age is $\sim 3$ to $4$ Gyr
in all three spectra. The STARLIGHT results show a wide spread in the 
Gyr-scale population ages, with quite well-defined very young 5-20 Myr 
populations in addition which tend to dominate the {\em light} except 
in the upper aperture. The STARLIGHT fits indicate that less than 
15\% of the light could come from populations younger than 25 Myr. 
Stellar mass is strongly dominated by the old 
population in all cases, with the oldest population in the lowest 
aperture similar to the Lick index results. Fitted mean metallicities 
are generally well below solar ($\sim 1/10$th of solar) for both young 
and old populations, except for a young component contributing to the 
lower spectrum, which is closer to solar. With the input models used 
with STARLIGHT we are not able to constrain abundance ratios in any way.

\begin{figure}
 \begin{center}
 \includegraphics[width=60mm, angle=-90]{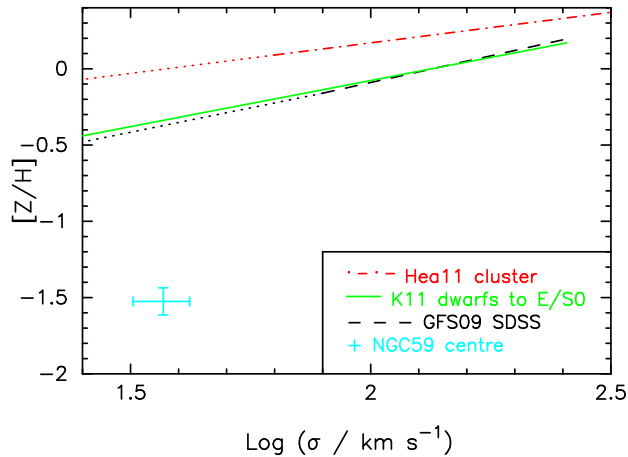} 
 \caption{ 
Plot showing [Z/H] relation
with Log($\sigma$), for samples of 
luminous early-type galaxies, compared with results from the 
current observations for NGC59. 
The mean trend from Graves et al. (2009) (GFS09) is shown in black 
and from Harrison et al. (2011) (Hea11) in red, with the current 
results for NGC59 highlighted by a cyan cross. 
The trend shown from Koleva et al (2011) in light green uses [Fe/H] 
as a proxy for metallicity and includes both dwarf and giant ETGs. 
Solid lines show the range of Log($\sigma$) covered by the references 
and extrapolations to lower Log($\sigma$) are shown by dotted lines.
}
\end{center}
\end{figure}

\subsection{Discussion} 

The results illustrated in Fig.~5 do not change significantly when other
indices combinations are selected, such as the higher order Balmer lines 
for ages, or the classic [Mg$<$Fe$>$] combination used by many authors as
an overall metallicity indicator. 
The left plot in Fig.~5 illustrates why it is difficult to constrain ages in
such low metallicity systems as NGC59, since the grid of models turns up to
higher Balmer line strengths, for a given age. This is less apparent in the
grid of models plotted in Fig.~6, since those models do not go down to such
low metallicities. 
The right panel in Fig.~5 illustrates the difficulty in
measuring [$\alpha$/Fe] accurately in such low metallicity systems, since 
the grids of SSP model predictions start to converge. This indicates how 
new measures of age and abundance patterns are needed in future, in order 
to understand the stellar population histories in these potential building 
block of luminous ETGs. There may be additional information in the bluer part 
of the spectrum, which is sensitive to abundance patterns (Sansom et al. 2013).

If low mass ETGs formed the building blocks of higher 
mass ETGs through dissipationless mergers then we might 
expect to see some relics of those building blocks still around today.
They are predicted to have the high metallicity, high [$\alpha$/Fe] ratios 
seen in luminous ETGs, if the build-up of galaxy mass is dissipationless. 
As yet no such low mass systems have 
been observed. The quasi-monolithic or early-hierarchical formation of 
ETGs easily accounts for enhaced [$\alpha$/Fe] ratios in more massive 
galaxies, since there was not time for Fe enrichment from delayed SNIa 
to feed back into the stellar population. However, simultaneously 
generating the higher metallicities in such a scenario is difficult 
(e.g. Pipino et al. 2009, their fig.~8; Merlin et al. 2012, their 
fig.~13). In fact NGC59 has unusually low metallicity for its mass, 
making it an unlikely candidate for a dissipationless building block.

A mass-metallicity plot is shown in Fig.~7, with stellar velocity dispersion 
used as a proxy for mass, in which NGC59 is contrasted with extrapolations
from higher $\sigma$ systems. The metallicity measured here for NGC59 is well 
below such extrapolations. These data for NGC59 also fall well below the 
[Fe/H] central measurements for dE/dS0 galaxy in the sample of 
Koleva et al. (2011), their fig.~13, at similar $\sigma$, with NGC59 
being near the low end of their $\sigma$ range. Koleva et al. (2011) found a 
range of, mostly negative, metallicity gradients in their dE/dS0 galaxies.

The above discussion is for metallicities of the stellar populations.
The gas metallicity measured in NGC59 is higher, typical of star-forming
dwarf galaxies of this mass, as shown in Section 3.2.
Systematic uncertainties in the emission line strengths arising from 
uncertainties in the best fit stellar template were found to be negligible.
Metallicities in HII gas associated with recent star formation will 
be higher than stellar abundances if the latest star formation is most 
metal enriched in a galaxy. However the estimated gas metallicity in NGC59 
is significantly higher than the stellar metallicity.
This may indicate that some or all of the gas was accreted from an 
enriched external source. Alternatively, simulations are needed to 
explore possible relationships between gas and star metallicities, 
for different star formation histories, to see how different 
the gas and star metallicities can become. 

More observations of statistically complete samples of lower mass 
ETGs are needed to probe their ranges of stellar population 
parameters and how the scaling relations behave at low masses. 
Such data will provide strong tests for hierarchical merger
models of galaxy evolution (e.g. Kaviraj et al. 2009). 
Future large telescopes, such as the E-ELT, will be able to resolve 
stellar populations in galaxies at the distance of the Sculptor group, 
which will allow both an independent check on their stellar metallicities 
and measurements of their stellar metallicity distributions.

\section[]{Conclusions}

Scaling relations for ETGs rely on accurate data down to 
low luminosities (e.g. the fundamental plane, or the mass-metallicity
relation discussed above). In this paper we looked at one such 
low luminosity ETG with data from the SALT telescope,
covering a broad wavelength range. The data stretch further into the 
blue region of the spectrum than most previous such studies, allowing 
for potentially stronger constraints on parameters such as stellar 
population age and abundance pattern (e.g. see Conroy, Graves \& 
van Dokkum 2014). We also looked across the centre of the system to 
see how the bulk of the stellar population behaves. 

Our main findings are that the low luminosity ETG, NGC59,
has a young (SSP age$\sim$ 2 to 6 Gyr), metal poor (SSP [Fe/H]$\sim -1.4$ 
to $-0.8$) population near its centre, with evidence of an older 
underlying population ($>\sim$ 8 Gyr) which is also very metal-poor. 
The [$\alpha$/Fe] ratio is difficult to accurately constrain with these data, 
for such a low metallicity galaxy, however, all measurements are consistent 
with slightly sub-solar values. Using two different sets of models led 
to similar results for the SSP parameters estimated from Lick line strengths. 
These results are qualitatively in agreement with our previous 
results for NGC59, but the measured metallicity is even lower than 
previously estimated. 
These data highlight the difficulty of measuring accurate ages for 
metal-poor systems, older than about 8 Gyrs.

Fitting two-burst SSP models gave a better representation of the 
overall spectrum and NGC59 was best fit with an old (14 Gyr), metal-poor 
SSP plus a 1.5 per cent mass frosting of young (0.07 Gyr), metal-poor stars.
Fits using STARLIGHT led to qualitatively similar results, confirming the 
low stellar metallicity in NGC59, although no abundance pattern constraints 
are yet possible with these full spectrum fitting methods.
We note that the gas phase metallicity in the HII region close to the 
nucleus of NGC59 is higher, [O/H]$\sim -0.7$, possibly suggesting external
gas accretion of unknown origin. 
Results such as the ones shown here for NGC59 and for larger samples 
of LLEs will be useful in future to constrain cosmological simulations 
and test hierarchical merging models of galaxy evolution and 
morphological transformations.

\section*{Acknowledgments}

All of the observations reported in this paper were obtained with the Southern
African Large Telescope (SALT). The proposal code was 2012-1-UKSC-003, and 
the PI was Sansom. Thanks go to the SALT support scientists for their help 
with these observations. PV and AYK acknowledge support from the National 
Research Foundation of South Africa. An RAS grant was awarded for an 
internship student, M.A.Deakin, to work on these data, plus a UClan 
internship for J.J. Thirlwall. Thanks to R. Proctor for the use of 
his SSP fitting software and to M. Sarzi for help with {\it pPXF} 
and {\it gandalf} software. We thank V.P. Debattista and the anonymous 
referee for helpful comments that improved the paper.

\bsp

\label{lastpage}

\end{document}